\documentclass[12pt]{article}
\usepackage{epsfig}
\usepackage{comment}
\usepackage{latexsym}
\usepackage{color}
\newcommand{\mysquare}[0]{\raise-.2ex\hbox{{\Large$\Box$}}}
\def\lsim{\mathrel{\rlap {\raise.5ex\hbox{$ < $}}
{\lower.5ex\hbox{$\sim$}}}}
\def\gsim{\mathrel{\rlap {\raise.5ex\hbox{$ > $}}
{\lower.5ex\hbox{$\sim$}}}} \topmargin -1.5cm \textheight=22.5cm \textwidth=16.5cm
\catcode`\@=11
\newcount\hour
\newcount\minute
\newtoks\amorpm
\hour=\time\divide\hour by60 \minute=\time{\multiply\hour by60 \global\advance\minute by-\hour}
\edef\standardtime{{\ifnum\hour<12 \global\amorpm={am}%
        \else\global\amorpm={pm}\advance\hour by-12 \fi
        \ifnum\hour=0 \hour=12 \fi
        \number\hour:\ifnum\minute<10 0\fi\number\minute\the\amorpm}}
\edef\militarytime{\number\hour:\ifnum\minute<10 0\fi\number\minute}
\def\draftlabel#1{{\@bsphack\if@filesw {\let\thepage\relax
   \xdef\@gtempa{\write\@auxout{\string
      \newlabel{#1}{{\@currentlabel}{\thepage}}}}}\@gtempa
   \if@nobreak \ifvmode\nobreak\fi\fi\fi\@esphack}
        \gdef\@eqnlabel{#1}}
\def\@eqnlabel{}
\def\@vacuum{}

\newcommand{\be}[0]{\begin{equation}}
\newcommand{\ee}[0]{\end{equation}}
\newcommand{\ba}[0]{\begin{eqnarray}}
\newcommand{\ea}[0]{\end{eqnarray}}

\def\bs{\begin{subequations}}
\def\es{\end{subequations}}

\def\thebibliography#1{%
\vskip 0.5cm \centerline{\bf \Large References}
\list{%
[\arabic{enumi}]}{\settowidth\labelwidth{[#1]} \leftmargin\labelwidth \advance\leftmargin\labelsep
\usecounter{enumi}}
\def\newblock{\hskip .11em plus .33em minus .07em}
\sloppy\clubpenalty4000\widowpenalty4000 \sfcode`\.=1000\relax}

\renewcommand{\section}{\setcounter{equation}{0}\@startsection
{section}{1}{0mm}{-\baselineskip}{0.5\baselineskip} {\normalfont\Large\bfseries}}

\renewcommand{\subsection}{\@startsection
{subsection}{2}{0mm}{-\baselineskip}{0.5\baselineskip} {\normalfont\large\bfseries}}

\renewcommand{\subsubsection}{\@startsection
{subsubsection}{3}{0mm}{-\baselineskip}{0.5\baselineskip} {\normalfont\normalsize\slshape}}

\usepackage{amssymb,amsfonts}
\usepackage{graphicx}
\usepackage{cite}

\def\bc{\begin{center}}
\def\ec{\end{center}}
\def\bea{\begin{eqnarray}}
\def\eea{\end{eqnarray}}

\def\F{{\cal F}}

\def\and{\quad\mbox{and}\quad}

\newcommand{\Z}{\mathbb{Z}}

\begin{document}
\begin{titlepage}
\begin{flushright}
LPTENS--08/44, 
May 2008
\end{flushright}

\vspace{2mm}

\begin{centering}
{\bf\huge Massive Boson-Fermion  Degeneracy\\}
$~$\\
{\bf \huge  and the} \\
{\bf\huge Early Structure of the Universe\\}

$~$\\
\vspace{.5 cm}
 {\Large \bf  Costas Kounnas}

\vspace{2mm}

 {\Large  Laboratoire de Physique Th\'eorique,\\
Ecole Normale Sup\'erieure, \\
24 rue Lhomond, F--75231 Paris cedex 05, France\\}
{\em  Costas.Kounnas@lpt.ens.fr}

\vspace {1.cm}

{\bf \Large Abstract}

\end{centering}
$~$\\
The existence of a new kind of {\it massive boson-fermion symmetry} is shown explicitly in the framework of the heterotic,  type II and type II orientifold superstring theories.  The target space-time is  two-dimensional. Higher dimensional models are defined via large marginal deformations of $J\bar J$-type. The  spectrum of the initial undeformed two dimensional vacuum consists of  massless boson degrees of freedom, while all massive  boson and fermion degrees of freedom exhibit a new Massive Spectrum Degeneracy Symmetry ($MSDS$) .  This precise property, distinguishes the $MSDS$ theories from the well known  supersymmetric $SUSY$-theories.  Some proposals are stated in the framework of these theories concerning the structure of:
$~~$\\
(i)  The Early Non-singular Phase of the Universe, \\
(ii)  The two dimensional boundary theory of $AdS_3$ Black-Holes, \\ 
(iii) Plausible  applications of the $MSDS$ theories in particle physics, alternative to 
 SUSY.
\vspace{5pt} \vfill \hrule width 6.7cm \vskip.1mm{\small \small \small \noindent Research
partially supported by the EU (under the contracts MRTN-CT-2004-005104, MRTN-CT-2004-512194,
MRTN-CT-2004-503369,  CNRS PICS 2530, 3059 and ANR (CNRS-USAR) contract
 05-BLAN-0079-01. Unit{\'e} mixte  du CNRS et de l'Ecole Normale Sup{\'e}rieure associ\'ee \`a
l'Universit\'e Pierre et Marie Curie (Paris
6), UMR 8549.}\\
\end{titlepage}
\newpage
\setcounter{footnote}{0}
\renewcommand{\thefootnote}{\arabic{footnote}}
 \setlength{\baselineskip}{.7cm} 

\setcounter{section}{0}
\section{Introduction}
In string theories the origin of space-time supersymmetry in 1+9 dimensions  follows from the  {\it equivalence}  of the vector-$V_8$ and spinor-$S_8$ of the  underlying  $SO(8)$ ``helicity" group  (light-cone picture)\cite{GSW}. In lower dimensions, depending on the choice of the internal compactified space, some of the space-time supersymmetries are broken. The  initial $SO(8)$ in 1+9 dimensions is  broken by  compactification to $SO(d-2)\times G$, (Helicity $\oplus$ $R$-symmetry group). The $R$-symmetry group $G$ characterizes the internal compactified space and ensures in lower space-time dimensions the number of the unbroken supersymmetries $N_d$ \cite{BD}, as well as a boson-fermion spectrum degeneracy. \\
$~$\\
In the string framework, $G$ is decomposed in left-and right- moving factors $G_L\times G_R$. It turns out that  the number of space-time supersymmtry depends on both $G_L$ and $G_R$ since both can induce  in space-time ``boson-fermion spectral flow degeneracies". In the world-sheet $G_L$  and $G_R$ are  expressed  in terms of  left- and right- moving  superconformal symmetries. I am stressing here the well-known correspondence between the number of space-time supersymmetries,  $N_d$, and the superconformal symmetries  ${\cal N}_{(L,R)}$  of the string world-sheet\cite{BD}, like for instance:

(i)  The existence of  $N_4=1$ supersymmetry in the heterotic, requiring $G_L=U(1)$ with  ${\cal N}_{(L,R)}=(2, 0)$ superconformal symmetry in the world-sheet \cite{BD} \cite{GEPN}.

(ii) When  $G_L=U(2)$ in the heterotic,  the supeconformal symmetry is at least ${\cal N}=(4,0)$ and the space-time supersymmetry becomes  $N_4=2$ .

(iii)  In  type II  theories supersymmetry emerges from left- and  right- moving sectors; then the space-time supersymmetry is : $N_4=1$ when ${\cal N}=(2,1)$; $N_4=2$  when ${\cal N}=(4,1)$ or ${\cal N}_{(L,R)}=(2,2)$;  $N_4=4$  when ${\cal N}_{(L,R)}=(4,4)$, ... .  
 
(iv) The maximal supersymmetry  ``in any type of superstring theory"  is achieved via toroidal compactifications  $T^n$, $n=10-d$,  leading to $N_4=4$ in heterotic or Type I and to $N_4=8$ in type IIA or IIB  superstrings.\\
$~$\\
 In our days, there are well known procedures  to reduce the number of supersymmetries, namely:  via symmetric orbifolds \cite{orbifolds} ($\equiv $ Calabi-Yau \cite{GSW}) compactification,  via fermionic \cite{ABK}\cite{KLT} and Gepner constructions\cite{GEPN},  asymmetric orbifolds\cite{ASorbifolds} ($\equiv$ generalized CY with torsion \cite{CYtorsion}, or type II orientifold compactifications \cite{Orientifolds} with or without  geometrical \cite{GeoFluxes} \cite{OpenFluxes} or non-geometrical \cite{Fluxes} fluxes. \\
$~$\\
The main achievement of the present work is the discovery  of a {\it new correspondence} between:\\
 $\bullet$ {\it A  Massive boson-fermion Spectrum-Degeneracy Symmetry in space-time, $ MSDS $ }\\
 $\bullet$ {\it  2d-wolrd-sheet Spectral-Flow Super-Conformal Symmetry  $ SFSC_{3\over 2}$}\\
  $~$\\
The initial construction assumes a two dimensional target space-time where the remaining extra eight internal dimensions are compact with a typical scale the string scale. The physical motivation of these 2-d constructions  is suggested  by the  ``Big-Bang" picture of general relativity that  predicts an initial classical singularity at the early times of the universe. Assuming a compact space, close to the singularity,  the typical scale of the universe reaches at early times  the  gravitational scale (string scale). Obviously at this early epoch the classical gravity is not valid any more and has to be replaced by a more fundamental singularity-free theory.\\
  $~$\\
 In the framework of superstrings, some  special  theories in $d\le 2$ exist, which are able to  describe successfully the early phase of the universe. These theories are non-singular and are based on a Spectral-Flow Super-Conformal Symmetry, $SFSC_{3\over 2}$, in the world-sheet. In the  space-time, the spectrum of the  massive  bosons and fermions  shows a Massive Spectrum Degeneracy Symmetry, $MSDS $.\\
  $~$\\
The other important point which is in favor to my proposal comes from the analysis of a phase transition above the Hagedorn temperature\cite{Hagedorn} $T\ge T_H$. It is well known that for high temperatures,  $T>T_H$, the string  partition function diverges due to the a thermal winding state which becomes tachyonic \cite{AtickWitten}\cite{RostKounnas}\cite{ AKADK}. This is a signal of a phase transition towards a new vacuum.  Many proposals were made in the literature concerning the high temperature non-singular phase\cite{AtickWitten},\cite{ AKADK} \cite{CosmoTopologyChange}. As I will argue in this work,  the  $MSDS$-vacua are  potential candidates, able to describe the early phase of the universe above $T_H$. 

\section{The maximally symmetric $MSDS$ Vacuum}
As was stated in the introduction, in the early stage of the universe all nine, or at least the eight- space coordinates are assumed to be compact and closed  to the string scale. Furthermore, the supersymmetric  vacuum cannot be the desire one since in any non-trivial cosmological  or thermal background the space-time supersymmetry is always broken.  The above two statements can be easily formulated in terms of the fermionic  construction\cite{ABK} \cite{KLT} where all 2d-world-sheet compact space coordinates are expressed in terms of free  2d-fermions. The advantage of this construction is due to a consistent separation of the left- and right-moving world-sheet degrees of freedom in terms of left- and right-moving 2d fermions which give us the possibility in manipulating easily the left--right asymmetric and even non-geometrical  constructions of vacua in string theory.

 \subsection{The type II  $MSDS$ Vacuum}
The starting example is in  the type II theories where  the left- and right-moving degrees of freedom are:\\
  $~$\\
$\bullet$ The light-cone degrees: $(\partial X^0,~\Psi^0)$, $(\partial X^L,~\Psi^L)$
  $~$\\
$\bullet$ The super-reparametrization ghosts: $(b,c)~(\beta,\gamma) $
  $~$\\
$\bullet$ The transverse super-coordinates $(\partial X^I,~\Psi^I),~I=1,...8$ .\\
 $~$\\
 In the fermionic construction the transverse super-coordinates are replaced by a set  of free fermions in the adjoint representation of a semi-simple gauge group  
 $H$\cite{ABKW} \cite{ABK}, $\{\chi^a\},~a=1,...n$, $~n={\rm dim } [\, H\,]=24$. The simplest choice of  $H$ is:
 \be
 H=SU(2)^8~\equiv~SO(4)^4,
 \ee
 where  the transverse super-coordinates, $(\partial X^I, ~\Psi^I)$ are replaced by
 $(y^I,w^I,\Psi_I)$ so that  for every $I=1,...,8$,  the coordinate currents $i\partial X^I =y^Iw^I$ are expressed in terms of $y^I,w^I$ 2d world-sheet fermions. For every $I$,  
 $\{y^I,w^I,\Psi_I\}$ define the adjoint representation, of $SU(2)_{k=2}$.
The choice $H= SU(2)^8$ of the coordinate-fermionisation is not unique. Other  non-trivial  choices of the coordinate-fermionisation exist based on different  $H$ :
 \be
 H=SU(5),~~ H=SO(7)\times SU(2),~~H=G_2\times Sp(4),~~H=SU(4)\times SU(2)^3,~~  H=SU(3)^3.
 \ee
In all above choices the dimension of $H$ is is always equal to 24, which as we will see it will be one  of the necessary condiions for the realization of the $MSDS$ symmetry. Once the choice of  boundary conditions on world-sheet respect the  global existence of the $H$ symmetry, then $H$ is promoted to a local gauge symmetry on the target  space-time\cite{ABKW}\cite{KF}. \\
  $~$\\
The fundamental operators are as usual the left- and right- moving energy-momentum tensor $T_B$ with conformal dimension $h_B=2$ and the superconformal operator $T_F$ with $h_F=3/2$ . $T_F$ is responsible for the local ${\rm \cal N}=1$ world-sheet superconformal symmetry \cite{GSW}.
 $$    
 T_B=-{1\over 2}(\partial X_0)^2 -{1\over 2}\Psi_0\partial \Psi_0+ {1\over 2}(\partial X_1)^2+{1\over 2}\Psi_1\partial \Psi_1~+~  \sum_{a=1}^{24}~{1\over 2}~\chi^a \partial  \chi^a 
 $$
 \be
T_F=i\partial X_0\Psi_0~+~i\partial X_1\Psi_1~+~f_{abc}~\chi^a  \chi^b\chi^c 
 \ee
where $f_{abc}$ are the structure constants of the group $H$. \\
  $~$\\
Following the rules of the fermionic construction and respecting the $H$-symmetry we can construct a very special tachyon free vacuum,  with left- right- holomorphic factorization of the partition function:  
  $$
  Z_{II}^{2d}=
\int_\F {d^2\tau\over ({\rm Im}\tau)^2}
 \left[{1\over 2} ~\sum_{a_L,b_L} (-)^{a_L+b_L}~~{\theta\left[^{a_L}_{b_L}\right]^{12 }\over \eta^{12}}\right]
 \left[{1\over 2}~\sum_{a_R,b_R } (-)^{a_R+b_R}~~{\bar\theta\left[^{a_R}_{b_R}\right]^{12 } \over { \bar \eta}^{12}}\right]  
  $$
   $~$\\
  or
  \be
    Z_{II}^{2d}=\int_\F {d^2\tau\over ({\rm Im}\tau)^2}
   \left [{(\theta_3^{12}-\theta_4^{12} )-  (\theta_2^{12}-\theta_1^{12}) \over 2\eta^{12}}\right] \left [{(\bar \theta_3^{12}-\bar \theta_4^{12} )-  (\bar \theta_2^{12}-\bar \theta_1^{12}) \over 2\bar \eta^{12}}\right]
  \ee
   $~$\\
  or in terms of the $SO(24)$ characters:
   \be
    Z_{II}^{2d}=\int_\F {d^2\tau\over ({\rm Im}\tau)^2}~~
   \left [~ V^{(24)}~-~S^{(24)} ~\right] \left [~\bar V^{(24)}~-~\bar S^{(24)} ~\right] 
    \ee
    The above expressions of $ Z_{II}^{2d}$ remain the same for any choice of left-and right-moving $H$-group, $H_L,~H_R$. In this respect {\it $ Z_{II}^{2d}$ is a unique tachyon free partition function}, (modulo the chirality of the left- and right-spinors), {\it that respects the $H_L\times H_R$ gauge symmetry}.\\
$~$\\    
 The holomorphic factorization of $ Z_{II}^{2d}$  reminds us the corresponding supersymmetric partition function in two dimensions obtained via $T^8$ compactification in two space-time dimensions with maximal supersymmetry:
  \be
    Z_{SUSY}^{2d}=\int_\F {d^2\tau\over ({\rm Im}\tau)^2} \left\{ {\Gamma_{(8,8)}\over {\eta^8\bar \eta^8}}\right\}
   \left [~ V^{(8)}~-~S^{(8)} ~\right] \left [~\bar V^{(8)}~-~\bar S^{(8)} ~\right] 
    \ee
 Modulo the non-holomorphic Narain lattice factor,  $\left\{ {\Gamma_{(8,8)}\over {\eta^8\bar \eta^8}}\right\}$ in $ Z_{SUSY}^{2d}$, the two partition functions share holomorphic factorization properties:\\
  $~$\\
 $\bullet$   $ Z_{SUSY}^{2d}$ is zero due to the equivalence of the bosonic  $V^{(8)}$  and  the fermionic $S^{(8)}$  $SO(8)$ characters.  This spectacular property
 follows from the $\theta$-function identities \cite{GSW}:
  \be
 \theta_3^{4}-\theta_4^{4} -  \theta_2^{4}=0, ~~~~~~ ~~~\theta_1^{4}=0\, ,
  \ee  
 having their  origin to the triality property of $SO(8)$ affine Lie algebra, namely, the equivalence of the vector $V^{(8)}$, the spinor $S^{(8)}$ and the conjugate spinor $C^{(8)}$ representations at any massive level.\\
  $~$\\
 $\bullet$ Contrary  to  SUSY  $ Z_{II}^{2d} $ is not zero. This is a signal of a non-equivalent  bosonic and fermionic mass spectrum. However, the situation is even more spectacular than the supersymmetric case! Thanks to the $SO(24)$ affine algebra\cite{theta12},  
 \be
  V^{(24)}~-~S^{(24)}={\rm constant} \equiv 24~
 \ee
 This follows from the theta function identities 
 $$
  \theta_3^{4}-\theta_4^{4} -  \theta_2^{4}=0, ~~~~\theta_1^{4}=0, ~~~~ \theta_2 \theta_3 \theta_4=2\eta^3
 $$
 that imply the identity 
 \be
 \label{theta12}
 {1\over 2} \left(~{\theta_3 ^{12} \over \eta ^{12}} ~-~ {\theta_4  ^{12} \over \eta ^{12}}~\right)~-~{1\over 2} \left(~{\theta_2 ^{12} \over \eta ^{12}}~-~{\theta_1 ^{12} \over \eta ^{12}}~\right)=24
 \ee
  $~$\\
 $\bullet$ The above identity shows that the spectrum of  massive bosons and massive fermions is identical to all string mass levels !  This is similar to the property of supersymmetic theories. In the massless level however the situation is different: Although  there are 24 left-moving   bosonic degrees of freedom there are no massless fermionic stares. Similarly there are 24 right-moving bosinic stares. In total there are $24\times 24=576$ scalar bosons at the  massless  level in the adjoint representation of $H_L$ and $H_R$ .\\
  $~$\\
 $\bullet$ The integrated partition function is thus  equal to $576\times {\rm \cal I} $, where ${\rm \cal I}$ is the integral over the fundamental domain
 $$
 {\rm \cal I}=\int_\F {d^2\tau\over ({\rm Im}\tau)^2}={\pi^2 \over 3},~~~~~~~~~ Z_H^{2d}={\pi^2 \over 3} \times 576 .
 $$
  $~$\\
 $\bullet$ 
 In SUSY models the underlying global gauge symmetry $H_L\times H_R$ is  necessarily   broken  to a discrete sub-group contrary to the  $MSDS$ model where $H_L\times H_R$ is unbroken, and is promoted to a local space-time gauge symmetry. \\
  $~$\\
In the next section I will examine in more detail the origin  of the boson-fermion  massive spectral symmetry, $MSDS$, and its difference from the  conventional supersymmetry.   

 \subsection{Chiral superconformal algebra and spectral flows in  $MSDS$}
 The symmetry operators of the $MSDS$ vacuum are the usual  holomorphic (anti-holomorphic  operators $T_B,~T_F$ ($\bar T_B,~\bar T_F$ ) giving rise to the standard $\rm {\cal N}=$ $(1,1)$ world-sheet superconformal symmetry induced by the familiar Operator Product Expansion \cite{GSW} (OPE) with $\hat c={2\over 3}c=8$.
$$
T_B (z)~T_B(w) ~{\sim}~  \frac{3\hat{c}}{4(z-w)^{4}}~+~\frac{2T_B(w)}{(z-w)^2}~+~
\frac{\partial T_B(w)}{(z-w)}
$$
\be
T_B (z)~ T_F(w) ~{\sim}~ \frac{3 T_F(w)}{2(z-w)^2}~+~
\frac{\partial T_F(w)}{(z-w)}\, ,~~~~
T_F(z)~T_F(w) ~ {\sim}~ \frac{\hat{c}}{(z-w)^3}~+~
\frac{2T_B(w)}{(z-w)}\nonumber \\
\ee
The extra symmetry operators are the $H$-currents:  $J^a\equiv f_{bc}^a \,\chi^b \chi^c$ and $\chi^a $ with conformal weights  $h_J=1$ and $h_{\chi}={1\over 2}$ respectively.
$$
T_B (z)~J^a(w) ~{\sim}~  \frac{J^a(w)}{(z-w)^2}~+~\frac{\partial J^a(w)}{(z-w)},~~~~~~~~~
T_B (z)~\chi^a(w) ~{\sim}~  \frac{\chi^a(w)}{2(z-w)^2}~+~\frac{\partial \chi^a(w)}{(z-w)},
$$
\be
\label{TJ}
T_F(z)~J^a(\omega)\sim {\chi^a \over (z-w)^2}~+~\frac{\partial \chi^a(w)}{(z-w)},~~~~~~~~~ T_F(z)~\chi^a(w)\sim {J^a \over (z-w)}~.~~~~~~~~~~
\ee
Finally there are two spin-field  operators with conformal weight ${3\over 2}$:
\be
C=Sp\{\chi^a\}_+ ~~~~~~~{\rm and}~~~~~~S=Sp\{\chi^a\}_-
\ee  
$$
T_B (z)~C(w) ~{\sim}~  \frac{3C(w)}{2(z-w)^2}~+~\frac{\partial C(w)}{(z-w)},~~~~~~T_B (z)~S(w) ~{\sim}~  \frac{3S(w)}{2(z-w)^2}~+~\frac{\partial S(w)}{(z-w)}
$$
\be
J^a(z)~C(w)_{\alpha}~{\sim}~f^a_{bc}~\sigma^{bc}_{\alpha\beta}~\frac{C(w)_{\beta}}{(z-w)}~,~~~~~~~~~~~~J^a(z)~S(w)_{\alpha}~{\sim}~f^a_{bc}~\sigma^{bc}_{\alpha\beta}~\frac{S(w)_{\beta}}{(z-w)}
\ee
where  $\sigma^{ab}_{\alpha\beta}$ are the $\sigma$-matrices of $SO(24)$:
$$
  \sigma^{ab}_{\alpha\beta}~=~{1\over 2}~[\gamma^a,\gamma^b]_{\alpha\beta}
$$
The indices $a,b$ are vector representation indices under $SO(24)$ and adjoint representation indices under $H$.  
$$
C(z)_{\alpha}~C(w)_{\beta}~{\sim}~{\delta_{\alpha\beta}\over (z-w)^3}~+~ { \sigma^{ab}_{\alpha\beta}~\chi_a\chi_b \over (z-w)^2}~+~{\delta_{\alpha\beta}~2T_B~+~(\gamma^a\gamma^b\gamma^c\gamma^d)_{\alpha\beta} ~(\chi_a\chi_b \chi_c\chi_d )\over (z-w)}~,~~~~~
$$
\be
S(z)_{\alpha}~S(w)_{\beta}~{\sim}~{\delta_{\alpha\beta}\over (z-w)^3}~+~ { \sigma^{ab}_{\alpha\beta}~\chi_a\chi_b \over (z-w)^2}~+~{\delta_{\alpha\beta}~2T_B~+~(\gamma^a\gamma^b\gamma^c\gamma^d)_{\alpha\beta} ~(\chi_a\chi_b \chi_c\chi_d )\over (z-w)}~.
\ee

\be
\label{CS}
C(z)_{\alpha}~S(w)_{\beta}~{\sim}~{1\over (z-w)^{1\over 2}}\left[{(\gamma^a)_{\alpha\beta}~\chi_a \over (z-w)^2}~+~{ (\gamma^a)_{\alpha\beta}~\partial \chi_a~+~(\gamma^a\gamma^b\gamma^c)_{\alpha\beta}~(\chi_a\chi_b\chi_c )\over (z-w)}\right]
\ee
  $~$\\
The above OPE relations between $T_B, T_F, C_{\alpha}, J^a, \chi_a $ define  {\it a new chiral superconformal algebra}.  As we will see in the next section, the $C(z)S(w)$ OPE in Eq. (\ref{CS}) implies a boson-fermion Spectral Flows  which guaranties the massive boson-fermion degeneracy of the Vacuum.

\subsection{Spectral flows and  the $MSDS$ operator-relations }
The vertex operators are dressed by the super-reparametrization ghosts \cite{FrShen}\cite{GSW}:  the space-time boson vertices are  expressed either in the zero or -1 ghost picture. The space-time fermions are in the $\pm {1\over2}$ picture.
\be
{\bf V}_b~=~ e^{-\Phi}~\chi_a,   {Ö}~~~~~~~{\bf S}_f~=~ e^{-{1\over 2}(\Phi + iH_0)}~S_{\alpha}
\ee
where the $H_0$ is the usual helicity field  defined via bosonization of $\Psi_0 $ and $\Psi_L$: $i\partial H_0 =\Psi_0 \Psi_L$. The conformal weight $h_{q}$ of the operator \cite{FrShen}\cite{GSW}
\be
e^{q\Phi} ~~\longrightarrow~~h_{q}=-\,{1\over 2}\,q\,(q+2)
\ee 
is such  that  ${\bf V}_b$ has conformal weight $h_b=1$ while ${\bf S}_f$ has a weight $h_f=2$. Thus, the string spectrum of bosons starts from a massless sector. Contrary, all space-time fermions are massive starting from the mass level 1. At the massless level, only  bosons are present: namely those which are created from the left- and right-moving vacuum  acting with the oscillators  $\chi^a_{-{1\over 2}}$ and $\bar\chi^a_{-{1\over2}}$ :
 \be
 ~\chi^a_{-{1\over 2}} ~|L>~\otimes~{\bar \chi}^a_{-{1\over 2}} ~ |R>
  \ee
At the massive level both bosons and fermions are present. At the first level we have the following states:
 \be
\left[\{\chi^a_{-{3\over2}}~\oplus~\chi^a_{-{1\over2}}~\chi^b_{-{1\over2}}~\chi^c_{-{1\over2}}\}\oplus  Sp\{\chi^a\}_- \right]|L> \otimes \left[\{\bar\chi^a_{-{3\over2}}\oplus\bar\chi^a_{-{1\over2}}~\bar\chi^b_{-{1\over2}}~\bar\chi^c_{-{1\over2}}\}\oplus Sp\{\bar\chi^a\}_- \right] |R> 
\ee 
where $Sp\{\chi^a\}_- $ is the spin field of $SO(n),~n=24$ with chirality ``$-$" and conformal weight, $h={n/16}={3/
 2}$. 
 
   $~$\\
Thanks to the identity \cite{theta12},
  \be
  \label{n24}
  n +{1\over 6}n(n-1)(n-2) ~=~ {1\over 2}~2^{n\over2}~=2048 ~~{\rm valid~ for}~~n=24
 \ee
the degeneracy of {\it space-time bosons}:
  $$
\{\chi^a_{-{3\over2}}~\oplus~\chi^a_{-{1\over2}}~\chi^b_{-{1\over2}}~\chi^c_{-{1\over2}}\} |L>~ \otimes ~\{\bar\chi^a_{-{3\over2}}\oplus\bar\chi^a_{-{1\over2}}~\bar\chi^b_{-{1\over2}}~\bar\chi^c_{-{1\over2}} \}|R> 
$$
\be
\oplus ~ Sp\{\chi^a\}_- |L> ~\otimes ~Sp\{\bar\chi^a\}_-  |R> 
\ee
and the degeneracy of {\it space-time fermions}: 
$$
  Sp\{\chi^a\}_-|L> ~\otimes ~\{\bar\chi^a_{-{3\over2}}\oplus\bar\chi^a_{-{1\over2}}~\bar\chi^b_{-{1\over2}}~\bar\chi^c_{-{1\over2}}\} |R> 
$$
 \be
\oplus ~ \{\chi^a_{-{3\over2}}~\oplus~\chi^a_{-{1\over2}}~\chi^b_{-{1\over2}}~\chi^c_{-{1\over2}}\}|L> ~\otimes ~Sp\{\bar\chi^a\}_-  |R> 
\ee 
are equal at least at the first massive level. This spectacular result is not a numerical  accident of the first  massive level but a deeper fundamental property of the affine $SO(24)$ algebra which leads to the $\theta^{12}$-identity in Eq. (\ref{theta12}).  The  relation at the first massive level in Eq. (\ref{n24})  together with the OPE's of the previous section indicate  the important role of the operator $O_{3/2}~$, ( $h_O=3/2$):
$$
O_{3/2}~\equiv ~(\gamma^a)_{\alpha\beta}~\partial \chi_a~+~(\gamma^a\gamma^b\gamma^c)_{\alpha\beta}\chi_a\chi_b\chi_c~\equiv~ \partial \hat \chi~+~\hat \chi~\hat \chi~ \hat \chi
$$
($\hat \chi$ is a short hand notation for $\gamma^a\chi_a$ where $\gamma^a$ are the $\gamma$-matrices of $SO(24)$ ). $O_{3/2}$ appears in the rhs of Eq. (\ref{CS}) and is used to define  a massive bosonic vertex operator in (-1) ghost  picture:
\be
{\bf B}_b= e^{-\Phi}~(~ \partial \hat \chi~+~\hat \chi~\hat \chi~ \hat \chi ~).
\ee
${\bf B}_b$  has conformal dimension $h_B=2$ and describes {\it massive bosonic states } starting at the first string level in a very similar way that ${\bf S}_f$ describes {\it  massive fermionic states} starting equally  at the first string level. The spectral flow of ${\bf B}$-states to ${\bf S}$-states
is express by the action of a ``Spectral Flow Operator"  ${\bf C}_{sf}$ :
$$
{\bf C}_{sf}=e^{{1\over2}(\Phi+iH_0)}~C~
$$
${\bf C}_{sf}$ is written in the (+1/2) ghost picture. It has conformal dimension $h_{\bf C}=1$ and $+1/2$ helicity charge. Thus, generically, 
 ${\bf C}_{sf}$ acting on ``physical"  bosonic states produces fermonic states (and vis versa). Although the ${\bf C}_{sf}$ action looks like a space-time  supersymmetry transformation, the actual  situation turns out to be drastically  different from that of supersymmetry. Indeed, the ${\bf C}_{sf}$ action leaves invariant the massless bosonic states of the theory, therefore the boson-to-fermion mapping does not exist for the massless states. This statement is visualized in the OPE :
 \be
 \label{CV}
 {\bf C}_{sf}(z)~{\bf V}_b(w)~\sim~ {\bf S}_f (w) \, ,~~{\rm finite~for ~} z\rightarrow  w
 \ee 
The absence of singular terms in ($z$-$w$) shows clearly that the massless states are invariant under ${\bf C}$-transformation. On the other hand the ${\bf C}$-action on the massive states is not-trivial:
\be 
\label{CB}
{\bf C}(z)~{\bf B}(w) ~\sim~{{\bf S} (w) \over (z-w) }~ +~{\rm finite~terms}
\ee
The above equation shows that the {\it massive bosonic states} are mapped to the fermionic ones.  To prove the reverse action  ${\bf C}(z) :~{\bf S}(w) \rightarrow {\bf B}(w)$ it  is not so direct since  standard picture changing  manipulation\cite{FrShen} are necessary to convert  ${\bf B}$ and ${\bf S}$ to their conventional ghost-pictures. This conversion utilizes   insertions of the ``picture changing operator", of conformal dimension zero, $e^{\phi}T_F$\cite{FrShen}\cite{BD},  in the OPE's. Modulo these manipulations, the
reverse action follows mainly from the Eq. (\ref{CS}) and the picture changing equivalence of $J^a=f^{a}_{bc} \chi^b\chi^c $  in the (0) ghost picture with the $e^{-\phi}\chi^a$ in the (-1) ghost picture (~see the OPE's in  Eqs (\ref{TJ}) ~).\\
$~$\\
 Summarizing: \\
$\bullet$    $T_B, T_F, C_{3/2}$ and $(J^a,\chi^a)$ define via the OPE's  a new super-conformal algebra. \\
$\bullet$  The closure of the algebra is  guarantied when $c=12$, so that $C_{3/2}$ is a  chirality   ``$+$"  spin-feld of $SO(24)$  with conformal weight $h_C=3/2$ . \\
$\bullet$  The realization of the algebra divides  the  ``physical"  states in two sectors :\\
 i) Massless sector which is invariant under  ${\bf C}$ spectral flow transformations.\\
 ii) Massive fermionic states ${\bf S}$ with  ``$-$"  chirality is  in one to one correspondence with the massive bosononic states.   ${\bf C}$ :  ${\bf S} \leftrightarrow {\bf B}$ $~~~~\rightarrow$ {\it massive supersymmetry}.
  
 \section{The heterotic  $MSDS$ Vacua}
 
In  the right-moving sector of the  heterotic string  the ${\rm \cal N}=1$ supersymmetry  is absent, while the left-moving degrees of freedom are as in the type II. The absence of the $2d$ conformal anomaly request $48$ right-moving world-sheet fermions. Following the basic rules of the fermionic construction and respecting the left-moving gauge  group $H_L$  one can construct at least three distinguishable vacuua  according to the gauge group of the right-moving sector:
  \be
  H_R= SO(48),~~~~~H_R=SO(32)\times E_8,~~~~H_R=E_8^3
  \ee
In all three cases   the heterotic partition function is tahyon free and  factorizes in  left- and right-holomorphic sectors:
 \be
  Z_{Het}^{2d}(H_R)=\int_\F {d^2\tau\over ({\rm Im}\tau)^2}
   \left [{(\theta_3^{12}-\theta_4^{12} )-  (\theta_2^{12}-\theta_1^{12}) \over 2\eta^{12}}\right] \bar \Gamma[H_R]
  \ee
where

$$
\bar \Gamma[SO(48)]={1\over2} \left [{(\theta_3^{24}+\theta_4^{24} )+ (\theta_2^{24}+\theta_1^{24}) \over \eta^{24}}\right]=1128+[~ j(\bar\tau)-744]
$$

$$
\bar \Gamma[SO(32)\times E_8]={1\over2} \left [{(\theta_3^{16}+\theta_4^{16} )+  (\theta_2^{16}+\theta_1^{16}) \over \eta^{16}}\right]\times {1\over2} \left [{(\theta_3^{8}+\theta_4^{8} )+  (\theta_2^{8}+\theta_1^{8}) \over \eta^{8}}\right]=744+[~j(\bar\tau)-744]
$$

 \be
 \bar \Gamma[E_8^3]=\left\{{1\over2} \left [{(\theta_3^{8}+\theta_4^{8} )+  (\theta_2^{8}+\theta_1^{8}) \over \eta^{8}}\right] \right\}^3=744+[~j(\bar\tau)-744]
  \ee
  $~$\\
  The three constructions give different number of massless states coming from the right-moving sector. The right-moving  massive states are expressed in terms of  the unique holomorphic modular invariant function  $j(\tau)$ of weight 24 . The  left-moving sector gives rise to $MSDS$ symmetry which implies a constant contribution $C_L=24$. Finally, the integrated partition function becomes:
  \be
   Z_{Het}^{2d}(G_R)={\pi^2\over 3}~C_L\times C_R
  \ee
  with
  $$
  C_L=24\, ,~~~~~C_R~[SO(32)\times E_8]~=~C_R~[E_8^3]=744\, ,~~~~~~C_R~[SO(48)]=1128\,.
  $$

 \subsection{The  type II Orientifold $MSDS$  Construction} 
Another interesting example with  $MSDS$ structure is the based to type II orientifolds.   
When $H_L=H_R$ the type II partition function becomes a perfect square.  This remark indicates  how to define the orientifold projection \cite{Orientifolds}
 $\Z_2$:
 \be
\Z_2~\equiv  \Omega.
 \ee
As usual  $\Omega~$ interchanges the left- and right- movers. 
Following the rules of orientifold construction, the torus partition function of the close string sector,  ${\cal T}= Z_H^{2d}~$ counts with the invariant linear sum of a state under $\Z_2$   and  as long as they are distinct. To complete the closed string states counting, one introduces the Klein bottle ${\cal K}$ in which only $\Z_2$ invariant states appear.
Each product of complex conjugate characters in ${\cal T}$ descend to a character in ${\cal K}$, with argument $2i\tau_2$. 
At the end the Klein bottle amplitude has a very simple form thanks to the $\theta^{12}$-identity:  
\be
{2\cal K}= \int_0^\infty \frac{d\tau_2}{\tau_2^2}\,~~
   \left[~ V^{(24)}~-~S^{(24)} ~\right] =\int_0^\infty \frac{d\tau_2}{\tau_2^2}\,~C~,~~{\rm with} ~~ C=24
   \ee
It turns out that the  annulus amplitude ${\cal A}$ has the same form modulo the Chan-Paton degeneracies: 
\be
{2\cal A}={n^2} ~ \int_0^\infty \frac{d\tau_2}{\tau_2^2}\,~~
   \left [~ V^{(24)}~-~S^{(24)} ~\right] =  {n^2} ~ \int_0^\infty \frac{d\tau_2}{\tau_2^2}\,~C~,~~{\rm with} ~~ C=24
\ee
The factor $n^2$ is introduced to account for  the Chan-Paton degeneracies.
Finally the Mobius amplitude  becomes:
\be
{2\cal M}=-{n} ~ \int_0^\infty \frac{d\tau_2}{\tau_2^2}\,~~
   \left [~ V^{(24)}~-~S^{(24)} ~\right] =-{n} ~ \int_0^\infty \frac{d\tau_2}{\tau_2^2}\ ~C~,~~{\rm with} ~~ C=24
\ee
As usual, the rank of the gauge group is determined by tadpole conditions for massless 
states, extracted from the transverse closed-string channel \cite{Orientifolds}. Clearly, this model is free of R-R 
tadpoles since  the R-R sectors are massive, while the 
dilaton tadpole reads \cite{Orientifolds},
\be
2\tilde K +2\tilde A+2\tilde M= \int_0^\infty {dl}~(2+{1\over 2}n^2-2n)~ C~=~0,~~~~{\rm for}~n=2.
\ee
The above expressions shows clearly that the absence of any tadpoles from  the open sector requires  the choice of $n=2$ and so,  the gauge group in the open sector is fixed to be  $SO(2)$. This situation is indeed extremely  interesting since it implies that the total contribution of the open sector and not only the tadpoles vanishes identically! This remarkable fact is again a concequence of the  $\theta^{12}$-identity which implies that the integrant $C$ is constant. Thanks to this fact, only the close sector gives non-trivial result in the partition function:
\be
Z^{2d}_{Orient}~=~{1\over 2}~Z_{II}^{2d}~
\ee

\section{Marginal deformation of the $MSDS$ vacua }  
 $~$\\ 
 In  type II  $MSDS$ vacua the  massless  bosons are $C_L\times C_R$  scalars.  $C_L$ and $C_R$ 
 are equal to the dimension of the adjoint representation of  gauge group $H_L$ and $H_R$ 
 respectively, ($C_L=C_R=24$).  These scalars parametrize  the manifold in a similar way like the 
 gauged supergravities: 
 \be
 {\rm\cal K}={SO(C_L,~C_R)\over SO(C_L)\times SO(C_R)}.
 \ee
 Because of the non-abelian structure of $H_L\times H_R$  the only marginal  deformations are those that correspond to the Cartan sub-algebra $U(1)^{r_L}\times U(1)^{r_R}$, with $r_L$ and $r_R$ is the rank of  $H_L$ and  $H_R$ respectively. The moduli space of these deformations, ($M_{IJ}~J_L^I  \times  J_R^{J}$ ), is reduced to:
 \be
 {\rm\cal M}={SO(r_L,~r_R)\over SO(r_L) \times SO(r_R)}~.
 \ee
 The maximal number of the moduli $M_{IJ}$ is when:
 \be 
 H_L= H_R ~=~SO(4)_{k=1}^4 ~=~ SU(2)_{k=2}^8 ,   ~~~~{\rm with} ~~~r_L=r_R=8\,.
 \ee
 In that precise case, the integrand of the deformed partition function factor out  {\it a shifted  lattice }
 $\Gamma_{8,8} (M)\left[^{h_i}_{g_i}\right] $; The latter couples non-trivially to the  ``parafermion numbers''  \cite{Parafermions}\cite{GEPN} defined by the gauge WZW-cosets \cite{GEPN}:
 $$
 \prod_{I_L=1,...8} \left( {SU(2)_{k=2}\over U(1)}\right)_{I_L}~\times \prod_{I_R=1,...8} \left( {SU(2)_{k=2}\over U(1)}\right)_{I_R} $$      
 The fact that the level $k$ is equal to $k=2$, makes the above coset structure to be  equivalent to eight left-moving world sheet fermions,  $\Psi_{I_L}$ and eight right-moving ones, $\Psi_{I_R} $.  At the end,  the shifted  $\Gamma_{8,8}$ lattice \cite{ShifftedLat} couples non-trivialy to the left-$\{ \Psi_{I_L}\}$ and right-$\{\Psi_{I_R} \}$ $R$-symmetry charges \cite{R-Shifted} of the conventional type II superstings ! Indeed, in the large moduli limit (modulo $S,T,U$-dualities), the  $\Gamma_{8,8}$ lattices decompactifies and the correlations with the $R$-symmetry charges become irrelevant \cite{R-Shifted}. At this limit one recovers the conventional ten dimensional type II supersymmetric vacua ! For large but not infinitely large deformations, the obtained vacua are those of  ``spontaneously broken supersymmetric vacua in the presence of geometrical fluxes'' \cite{ GeoFluxes} studied in details  refs \cite{ R-Shifted}\cite{ Cosmo-RT-Shifted}. Furthermore, some of the Euclidian  version of the models, correspond to ``thermal stringy vacua" in the presence of non-trivial left-right asymmetric  ``gravito-magnetic fluxes" studied recently in refs \cite{Cosmo-AS-RT-Shifted}\cite{Cosmo-RT-Shifted}.  The would be ``initial" classical singularity of general relativity as well as the stringy Hagedorn-like singularities are both resolved by these fluxes ! \cite{Cosmo-AS-RT-Shifted}\\
 $~$\\
 The above generic properties of the deformed $MSDS$ vacua, strongly  suggest the following \\
  {\it \bf Cosmological Conjecture}:
 \begin{itemize}
 \item
  The $MSDS$ vacua, (or even less symmetric orbifold versions of those), are potential candidates  able to describe {\it the early non-singular phase of a stringy cosmological universe}. 
  \item 
  During the cosmological evolution the deformation moduli $M_{IJ}\rightarrow M_{IJ}(t)$ evolves with the time. Once  $M_{IJ}(t)$ are  sufficiently large (modulo $S,T,U$-dualities), an effective field theory description  emerges with an induced ``space-time geometry" of an {\it effective higher dimensional  space-time}. Also, the relevant degrees of freedom and interactions are well described by some``no-scale" effective supergravity  theories\cite{Noscale} of the  conventional superstrings \cite{StringyNoscale}.
  \item
  The effects of the initial  $MSDS$ structure induces at the large moduli limit non-trivial ``geometrical" fluxes \cite{GeoFluxes},\cite{ R-Shifted}\cite{Cosmo-AS-RT-Shifted} which  in the language of the effective supergravity give rise to a spontaneous breaking of supersymmetry \cite{GeoFluxes}\cite{AKADK} and to a finite temperature description of the effective theory \cite{AKADK} \cite{Cosmo-RT-Shifted} \cite{Cosmo-AS-RT-Shifted}
 \end{itemize}
 $~$\\ 
 The above discussion still valid for the other choices of $H_L,H_R$. The rank however is reduces:\\
(i)  $~r=4~$ for $H=SU(5)$,  $~H=SO(7)\times SU(2)$ or $H=G_2\times Sp(4)$\, ,\\
(ii) $r=6~$ for $H=SU(4)\times SU(2)^3$  or $H=SU(3)^3$.\\
 In the large $M_{IJ}$-moduli limit of the deformed $H=SU(2)^8$ vacua, the effective geometric description would be up to10-dimensional space-time. In the lower rank cases however, the effective space-time cannot be higher than 6-dimensional in the case (i) and not higher than 8-dimensional in the case (ii). \\
$~$\\
The Heterotic $MSDS$ vacua contain  much more massless states. Here $C_R=744$ when $H_R=SO(32)\times E_8$ or $H=E_8^3$ and $C_R=1128$ when $H_R=SO(48)$. In all cases  however the dimension of the right-moving gauge group is $C_R=24$. The previous properties of the deformed  $MSDS$ type II vacua, still valid in the heterotic and in the orientifold deformed $MSDS$ vacua. \\

\section{Further perspectives and conclusions} 
The existence of a new of massive boson-fermion degeneracy symmetry, is shown by explicit construction  in the framework of Heterotic, Type II and Type II orientifold superstring theories. 
In all constructions  the target space-time is 2-dimensional and their spectrum consist of massless bosonic degrees of freedom while all massive boson and fermion degrees of freedom show a new Massive Spectrum Degeneracy Symmetry $MSDS$. This remarkable property follows from the
modular properties between the Vector ($V$), Spinor ($S$) and Anti-Spinor ($C$) characters of the affine $SO(24)$ algebra that are formulated algebraically in terms of $\theta^{12}$-identity, Eq. (\ref{theta12}).\\
$~$\\
A new chiral $N=1$ superconformal algebra is proposed based on the usual $N=1$ super-Virassoro operators,  $T_B,~(h_B=2)$ and  $T_F, ~(h_F=3/2)$ together with  $C, ~(h_C=3/2)$ and $J^a, ~(h_J=1)$, with $J^a$ are the currents of a semi-simple gauge group $H$, of dimension $d_H=24$. A simple realization of this new algebra is given in terms of world-sheet fermions. The massive boson-fermion degeneracy follows from a ``spectral flow" relations induced by the algebra $\{ T_B,~T_F, ~C, ~J^a \} $. \\
$~$\\
In all $MSDS$-vacua, the non-abelian gauge group $H$ is unbroken. Marginal deformations of current-current type, $M_{IJ}~J^I_L\times J_R$) breaks {\it spontaneously} $H_L$ and $H_R$ to abelian sub-groups $U(1)^{r_L}\times   U(1)^{r_R}$. What is extremely  interesting is the fact that in the {\it large $M_{IJ}$ deformation limit}  the strongly-deformed $MSDS$-vacua, are well described in terms of  an effective  ``higher dimensional" conventional  superstring theory in which the space-time supersymmetry is {\it spontaneously broken}  by ``geometrical" and ``thermal" fluxes. This fundamental generic  properties of the  deformed $MSDS$-vacua, strongly  suggest to consider them as the most (semi-) realistic candidates able to describe the  ``early non-singular phase of our Universe", free of initial ``general relativity singularity" and free of any ``Hagedorn-like" stringy singularities. Further investigations in this direction are necessary, and will be exhaustively studied by the author and collaborators in future publications. \\
$~$\\
Another noticeable  property of the 2-d $MSDS$-vacua is the {\it holomorphic  factorization} property of their partition function.  Although these theories have non trivial massive spectrum, thanks to $MSDS$, all non-topological degrees of freedom are effectively washed out from the partition function ! In that respect, $MSDS$-vacua realize  2-d target-space conformal field theories with the holomorphic factorization properties similar to those initially proposed by Witten \cite{WittenHoloFactor} in the context  of BTZ black holes \cite{HennStrominger}. In this context   2-d   $MSDS$-vacua (mainly the heterotic ones), are identified with the boundary 2-d conformal field theory of $AdS_3$ \cite{HennStrominger}. Following Witten's conjecture, the massive  bosonic spectrum is identified with the mass spectrum of BTZ-black-holes \cite{WittenHoloFactor}. The $MSDS$ theories however, suggest more, mainly,  the existence of a ``massive supersymmetry"  fermionic partner having the same mass spectrum as the bosonic one ! \\
 $~$\\
 The other interesting structure of $MSDS$-vacua, is their connection to the ``gauge supergravity theories".  Although this connection is {\bf not yet }  well transparent in the undeformed $MSDS$-vacua, (where the description is non-geometrical),  it is well established however  in the ``strongly deformed phase" via the induced geometrical fluxes of the effective higher dimensional theories.\\
 $~$\\
 Finally, an interesting question is whether it is possible to construct in a higher than two dimensions
 a field theory with an unbroken $MSDS$, and in particular in $ d=4$.  A progress in that direction may give an alternative to the conventional supersymmetry approach concerning the well known mathematical inconsistencies related to the hierarchy and to the cosmological constant problem.   
 

\section*{Acknowledgements}

I am grateful to Luis Alvarez-Gaum\'e, Constantin Bachas,  David Gross, Andr\'e Neuv\'e, Herman Nicolai, Herv\'e Partouche, Pierre Ramond and especially to Carlo Angelantonj, Nicolaos Toumbas  and Jan Troost for useful and fruitful  discussions.
This work  is partially supported by the EU contract MRTN-CT-2004-005104
and the ANR (CNRS-USAR) contract  05-BLAN-0079-01.



\begin{thebibliography}{99}

\bibitem{GSW}
  M.~B.~Green, J.~H.~Schwarz and E.~Witten,
  ``Superstring Theory", Vol. 1 and Vol. 2:
{\it  Cambridge, UK: Univ. Pr.} ( 1987), Cambridge Monographs on Mathematical Physics.

  J.~Polchinski,
  ``String theory" Vol. 1 and Vol. 2: ``An introduction to the bosonic string,'' and 
``Superstring theory and beyond,''
{\it  Cambridge, UK: Univ. Pr. (1998) 402 p} and {\it  Cambridge, UK: Univ. Pr. (1998) 531 p}.


  E.~Kiritsis,
  ``Introduction to superstring theory,'' {\it Leuven Notes in Mathematical and Theoretical Physics}, V.9: arXiv:hep-th/9709062,
  E.~Kiritsis,
  ``String theory in a nutshell,''
{\it  Princeton, USA: Univ. Pr. (2007) 588 p}.


\bibitem{BD}
  T.~Banks and L.~J.~Dixon,
  ``Constraints on String Vacua with Space-Time Supersymmetry,''
  Nucl.\ Phys.\  B {\bf 307} (1988) 93.
  


\bibitem{GEPN}  
  D.~Gepner and Z.~A.~Qiu,
  ``Modular Invariant Partition Functions for Parafermionic Field Theories,''
  Nucl.\ Phys.\  B {\bf 285}, 423 (1987).
  

  D.~Gepner,
  ``On the Spectrum of 2D Conformal Field Theories,''
  Nucl.\ Phys.\  B {\bf 287}, 111 (1987).


  D.~Gepner,
  ``Space-Time Supersymmetry in Compactified String Theory
 and Superconformal Models,''
  Nucl.\ Phys.\  B {\bf 296}, 757 (1988).
 
  D.~Gepner,
  ``On the algebraic structure of N=2 string theory,''
  Commun.\ Math.\ Phys.\  {\bf 142}, 433 (1991).
  

\bibitem{orbifolds}
  L.~J.~Dixon, J.~A.~Harvey, C.~Vafa and E.~Witten,
  ``Strings on Orbifolds,''
  Nucl.\ Phys.\  B {\bf 261}, 678 (1985).
  

  L.~J.~Dixon, J.~A.~Harvey, C.~Vafa and E.~Witten,
  ``Strings on Orbifolds. 2,''
  Nucl.\ Phys.\  B {\bf 274}, 285 (1986).
 



\bibitem{ABK}
  I.~Antoniadis, C.~P.~Bachas and C.~Kounnas,
  ``Four-Dimensional Superstrings,''
  Nucl.\ Phys.\  B {\bf 289} (1987) 87.
  


\bibitem{KLT}
  H.~Kawai, D.~C.~Lewellen and S.~H.~H.~Tye,
 ``Construction of Fermionic String Models in Four-Dimensions,''
  Nucl.\ Phys.\  B {\bf 288}, 1 (1987).
  


\bibitem{ASorbifolds}
  K.~S.~Narain, M.~H.~Sarmadi and C.~Vafa,
  ``Asymmetric Orbifolds,''
  Nucl.\ Phys.\  B {\bf 288}, 551 (1987).
  

  K.~S.~Narain, M.~H.~Sarmadi and C.~Vafa,
  ``Asymmetric orbifolds: Path integral and operator formulations,''
  Nucl.\ Phys.\  B {\bf 356}, 163 (1991).
 

\bibitem{Orientifolds}
  C.~Angelantonj and A.~Sagnotti,
  ``Open strings,''
  Phys.\ Rept.\  {\bf 371} (2002) 1
  [Erratum-ibid.\  {\bf 376} (2003) 339]
  [arXiv:hep-th/0204089].
  

\bibitem{CYtorsion}
M.~Grana, T.~W.~Grimm, H.~Jockers and J.~Louis,
  ``Soft supersymmetry breaking in Calabi-Yau orientifolds with D-branes  and
  fluxes,''
  Nucl.\ Phys.\  B {\bf 690} (2004) 21
  [arXiv:hep-th/0312232].
  
 D.~Lust, S.~Reffert and S.~Stieberger,
  ``Flux-induced soft supersymmetry breaking in chiral type IIb  orientifolds
  with D3/D7-branes,''
  Nucl.\ Phys.\  B {\bf 706} (2005) 3
  [arXiv:hep-th/0406092].
 


\bibitem{GeoFluxes}
J.~P.~Derendinger, C.~Kounnas, P.~M.~Petropoulos and F.~Zwirner,
  ``Superpotentials in IIA compactifications with general fluxes,''
  Nucl.\ Phys.\  B {\bf 715} (2005) 211
  [arXiv:hep-th/0411276].

  
  J.~P.~Derendinger, C.~Kounnas, P.~M.~Petropoulos and F.~Zwirner,
  ``Fluxes and gaugings: $N$ = 1 effective superpotentials,''
  Fortsch.\ Phys.\  {\bf 53} (2005) 926
  [arXiv:hep-th/0503229].


  G.~Villadoro and F.~Zwirner,
  ``N = 1 effective potential from dual type-IIA D6/O6 orientifolds with
  general fluxes,''
  JHEP {\bf 0506}, 047 (2005)
  [arXiv:hep-th/0503169].

  G.~Villadoro and F.~Zwirner,
  ``D terms from D-branes, gauge invariance and moduli stabilization in  flux
  compactifications,''
  JHEP {\bf 0603}, 087 (2006)
  [arXiv:hep-th/0602120].

L.~Andrianopoli, M.~A.~Lledo and M.~Trigiante,
  ``The Scherk-Schwarz mechanism as a flux compactification with internal
  torsion,''
  JHEP {\bf 0505} (2005) 051
  [arXiv:hep-th/0502083].
  
G.~Dall'Agata and N.~Prezas,
  ``Scherk-Schwarz reduction of M-theory on $G_2$-manifolds with fluxes,''
  JHEP {\bf 0510} (2005) 103
  [arXiv:hep-th/0509052].
 

  J.~P.~Derendinger, P.~M.~Petropoulos and N.~Prezas,
  ``Axionic symmetry gaugings in N = 4 supergravities and their
  higher-dimensional origin,''
  Nucl.\ Phys.\  B {\bf 785}, 115 (2007)
  [arXiv:0705.0008 [hep-th]].




 \bibitem{OpenFluxes}
   
  C.~Angelantonj, S.~Ferrara and M.~Trigiante,
  ``New D = 4 gauged supergravities from N = 4 orientifolds with fluxes,''
  JHEP {\bf 0310}, 015 (2003)
  [arXiv:hep-th/0306185].

  
  C.~Angelantonj, R.~D'Auria, S.~Ferrara and M.~Trigiante,
  ``K3 x T**2/Z(2) orientifolds with fluxes, open string moduli and  critical
  Phys.\ Lett.\  B {\bf 583}, 331 (2004)
  [arXiv:hep-th/0312019].

  C.~Angelantonj, S.~Ferrara and M.~Trigiante,
  ``Unusual gauged supergravities from type IIA and type IIB orientifolds,''
  Phys.\ Lett.\  B {\bf 582}, 263 (2004)
  [arXiv:hep-th/0310136].

 C.~Angelantonj, R.~D'Auria, S.~Ferrara and M.~Trigiante,
  ``$K3 \times T^2/Z_2$ orientifolds with fluxes, open string moduli and  critical points,''
  Phys.\ Lett.\  B {\bf 583} (2004) 331
  [arXiv:hep-th/0312019].

  



\bibitem{Fluxes}
  K.~Dasgupta, G.~Rajesh and S.~Sethi,
  ``M theory, orientifolds and G-flux,''
  JHEP {\bf 9908} (1999) 023
  [arXiv:hep-th/9908088].
 

  
  A.~R.~Frey and J.~Polchinski,
  ``N = 3 warped compactifications,''
  Phys.\ Rev.\  D {\bf 65} (2002) 126009
  [arXiv:hep-th/0201029].


  S.~Gukov, C.~Vafa and E.~Witten,
  ``CFT's from Calabi-Yau four-folds,''
  Nucl.\ Phys.\  B {\bf 584} (2000) 69
  [Erratum-ibid.\  B {\bf 608} (2001) 477]
  [arXiv:hep-th/9906070].
  
    
  S.~B.~Giddings, S.~Kachru and J.~Polchinski,
  ``Hierarchies from fluxes in string compactifications,''
  Phys.\ Rev.\  D {\bf 66}, 106006 (2002)
  [arXiv:hep-th/0105097].

  


\bibitem{Hagedorn}
  R.~Hagedorn,
 ``Statistical thermodynamics of strong interactions at high-energies,''
  Nuovo Cim.\ Suppl.\  {\bf 3}, 147 (1965).

  S.~Fubini and G.~Veneziano,
  ``Level structure of dual-resonance models,''
  Nuovo Cim.\  A {\bf 64}, 811 (1969).
  

  K.~Bardakci and S.~Mandelstam,
  ``Analytic solution of the linear-trajectory bootstrap,''
  Phys.\ Rev.\  {\bf 184}, 1640 (1969).
 

  K.~Huang and S.~Weinberg,
  ``Ultimate temperature and the early universe,''
  Phys.\ Rev.\ Lett.\  {\bf 25}, 895 (1970).
  

  B.~Sathiapalan,
  ``Vortices on the String World Sheet and Constraints on Toral
  Compactification,''
  Phys.\ Rev.\  D {\bf 35}, 3277 (1987).
  

  Y.~I.~Kogan,
  ``Vortices On The World Sheet And String's Critical Dynamics,''
  JETP Lett.\  {\bf 45}, 709 (1987)
  [Pisma Zh.\ Eksp.\ Teor.\ Fiz.\  {\bf 45}, 556 (1987)].
  

  M.~Axenides, S.~D.~Ellis and C.~Kounnas,
 ``Universal Behavior Of D-Dimensional Superstring Models,''
  Phys.\ Rev.\  D {\bf 37}, 2964 (1988).
 

  R.~H.~Brandenberger and C.~Vafa,
  ``Superstrings in the Early Universe,''
  Nucl.\ Phys.\  B {\bf 316} (1989) 391.
 
 
  
 A.~Tseytlin and C.~Vafa,
  ``Elements of String Cosmology,''
  Nucl.\ Phys.\  B {\bf 372} (1992) 443
  [arXiv:hep-th/9109048].


  D.~Kutasov and N.~Seiberg,
  ``Number Of Degrees Of Freedom, Density Of States And Tachyons In String
  Theory And Cft,''
  Nucl.\ Phys.\  B {\bf 358} (1991) 600.

 
  D.~Israel and V.~Niarchos,
  ``Tree-level stability without spacetime fermions: Novel examples in string
  theory,''
  JHEP {\bf 0707}, 065 (2007)
  [arXiv:0705.2140 [hep-th]].



  
\bibitem{AtickWitten}
  J.~Atick and E.~Witten,
  ``The Hagedorn Transition and the Number of Degrees of Freedom of String
  Theory,''
  Nucl.\ Phys.\  B {\bf 310}, 291 (1988).
  


\bibitem{RostKounnas}
  C.~Kounnas and B.~Rostand,
 ``Coordinate Dependent Compactifications and Discrete Symmetries,''
  Nucl.\ Phys.\  B {\bf 341} (1990) 641.
  




\bibitem {AKADK}
  I.~Antoniadis and C.~Kounnas,
  ``Superstring phase transition at high temperature,''
  Phys.\ Lett.\  B {\bf 261} (1991) 369.
  

  I.~Antoniadis, J.~P.~Derendinger and C.~Kounnas,
  ``Non-perturbative supersymmetry breaking and finite temperature
  instabilities in  $N$ = 4 superstrings,''
  arXiv:hep-th/9908137.

  C.~Kounnas,
 ``Universal thermal instabilities and the high-temperature phase of the  $N$ =
  4 superstrings,''
  arXiv:hep-th/9902072.
 
 
  
\bibitem{CosmoTopologyChange}  
  E.~Kiritsis and C.~Kounnas,
  ``Dynamical topology change, compactification and waves in a stringy early
  universe,''
  arXiv:hep-th/9407005.


  
  E.~Kiritsis and C.~Kounnas,
  ``Dynamical topology change in string theory,''
  Phys.\ Lett.\  B {\bf 331} (1994) 51
  [arXiv:hep-th/9404092].

  E.~Kiritsis and C.~Kounnas,
  ``Dynamical topology change, compactification and waves in string
  cosmology,''
  Nucl.\ Phys.\ Proc.\ Suppl.\  {\bf 41} (1995) 311
  [arXiv:gr-qc/9701005].

  E.~Kiritsis and C.~Kounnas,
  ``String gravity and cosmology: Some new ideas,''
  arXiv:gr-qc/9509017.


 

\bibitem{ABKW}
  I.~Antoniadis, C.~Bachas, C.~Kounnas and P.~Windey,
  ``Supersymmetry among free fermions and superstrings,''
  Phys.\ Lett.\  B {\bf 171}, 51 (1986).
  


\bibitem{KF}
  S.~Ferrara and C.~Kounnas,
  ``Extended Supersymmetry In Four-Dimensional Type II Strings,''
  Nucl.\ Phys.\  B {\bf 328} (1989) 406.
  


\bibitem{theta12}
  C.~Kounnas, B.~Rostand and E.~T.~Tomboulis,
  ``Heterotic (2,1) supergravity in two-dimensions,''
  Nucl.\ Phys.\  B {\bf 359} (1991) 673.
  


\bibitem{FrShen}
  D.~Friedan, S.~H.~Shenker and E.~J.~Martinec,
  ``Covariant Quantization Of Superstrings,''
  Phys.\ Lett.\  B {\bf 160}, 55 (1985).

  D.~Friedan, E.~J.~Martinec and S.~H.~Shenker,
  ``Conformal Invariance, Supersymmetry And String Theory,''
  Nucl.\ Phys.\  B {\bf 271}, 93 (1986).

  J.~Cohn, D.~Friedan, Z.~A.~Qiu and S.~H.~Shenker,
  ``Covariant Quantization of Supersymmetric string theories"
  Nucl.\ Phys.\  B {\bf 278}, 577 (1986).
  
  
  
\bibitem{Parafermions}
  V.~A.~Fateev and A.~B.~Zamolodchikov,
  ``Parafermionic Currents In The Two-Dimensional Conformal Quantum Field
  Theory And Selfdual Critical Points In Z(N) Invariant Statistical Systems,''
  Sov.\ Phys.\ JETP {\bf 62}, 215 (1985)
  [Zh.\ Eksp.\ Teor.\ Fiz.\  {\bf 89}, 380 (1985)].
  

  A.~B.~Zamolodchikov and V.~A.~Fateev,
  ``Disorder Fields in Two-Dimensional Conformal Quantum Field Theory and N=2
  Extended Supersymmetry,''
  Sov.\ Phys.\ JETP {\bf 63}, 913 (1986)
  [Zh.\ Eksp.\ Teor.\ Fiz.\  {\bf 90}, 1553 (1986)].

  Z.~A~Qiu,
  ``Nonlocal current algebra and N=2 superconformal field theory in two dimensions"
  Phys.\ Lett.\  B {\bf 188}, 207 (1987).
  
  V.~G.~Kac and D.~H.~Peterson,
  ``Infinite dimensional Lie algebras, theta functions and modular forms,''
  Adv.\ Math.\  {\bf 53}, 125 (1984).
  


\bibitem{ShifftedLat}
  C.~Kounnas,
  ``BPS states in superstrings with spontaneously broken SUSY,''
  Nucl.\ Phys.\ Proc.\ Suppl.\  {\bf 58} (1997) 57
  [arXiv:hep-th/9703198].
  
 
  E.~Kiritsis and C.~Kounnas,
  ``Perturbative and non-perturbative partial supersymmetry breaking: $ N$ = 4
$  \rightarrow N $= 2 $\rightarrow N$ = 1,''
  Nucl.\ Phys.\  B {\bf 503} (1997) 117
  [arXiv:hep-th/9703059].
  

  E.~Kiritsis, C.~Kounnas, P.~M.~Petropoulos and J.~Rizos,
  ``String threshold corrections in models with spontaneously broken
  supersymmetry,''
  Nucl.\ Phys.\  B {\bf 540} (1999) 87
  [arXiv:hep-th/9807067].


\bibitem{R-Shifted}
  J.~Scherk and J.~H.~Schwarz,
  ``Spontaneous breaking of supersymmetry through dimensional reduction,''
  Phys.\ Lett.\  B {\bf 82} (1979) 60.

  R.~Rohm,
  ``Spontaneous supersymmetry breaking in supersymmetric string theories,''
  Nucl.\ Phys.\  B {\bf 237} (1984) 553.
  
  C.~Kounnas and M.~Porrati,
  ``Spontaneous Supersymmetry Breaking in String Theory,''
  Nucl.\ Phys.\  B {\bf 310} (1988) 355.
  
  
  S.~Ferrara, C.~Kounnas and M.~Porrati,
  ``N=1 Superstrings With Spontaneously Broken Symmetries,''
  Phys.\ Lett.\  B {\bf 206} (1988) 25.

  
  
  S.~Ferrara, C.~Kounnas, M.~Porrati and F.~Zwirner,
  ``Effective Superhiggs and Strm**2 from Four-dimensional Strings,''
  Phys.\ Lett.\  B {\bf 194} (1987) 366.


  S.~Ferrara, C.~Kounnas and M.~Porrati,
``Superstring Solutions with Spontaneously Broken Four-Dimensional
  Supersymmetry,''
  Nucl.\ Phys.\  B {\bf 304} (1988) 500.
 



\bibitem{Cosmo-RT-Shifted}
  C.~Kounnas, N.~Toumbas and J.~Troost,
 ``A Wave-function for Stringy Universes,''
  JHEP {\bf 0708} (2007) 018
  arXiv:0704.1996 [hep-th].
 

  T.~Catelin-Jullien, C.~Kounnas, H.~Partouche and N.~Toumbas,
  ``Thermal/quantum effects and induced superstring cosmologies,''
  Nucl.\ Phys.\  B {\bf 797} (2008) 137
  arXiv:0710.3895 [hep-th].
 

  T.~Catelin-Jullien, C.~Kounnas, H.~Partouche and N.~Toumbas,
 ``Thermal and quantum superstring cosmologies,''
  arXiv:0803.2674 [hep-th].
  

T.~Catelin-Jullien, C.~Kounnas, H.~Partouche and N.~Toumbas,
``Induced superstring  cosmologies and moduli stabilization,"
LPTENS--08/42, CPHT--RR054.0708 preprint.



\bibitem{Cosmo-AS-RT-Shifted}
  C.~Angelantonj, M.~Cardella and N.~Irges,
  ``An alternative for moduli stabilisation,''
  Phys.\ Lett.\  B {\bf 641} (2006) 474
  [arXiv:hep-th/0608022].
  
C. ~Angelantonj, C.~Kounnas,
 H.~Partouche and N.~Toumbas
 ``Resolution of Hagedorn singularity in superstrings with gravito-magnetic flaxes" ,
LPTENS--08/43,  CPHT--RR055.0708,  [arXiv:0808.1357 [hep-th]].


\bibitem{Noscale}
  E.~Cremmer, S.~Ferrara, C.~Kounnas and D.~V.~Nanopoulos,
  ``Naturally Vanishing Cosmological Constant In N=1 Supergravity,''
  Phys.\ Lett.\  B {\bf 133} (1983) 61.
 
  J.~R.~Ellis, C.~Kounnas and D.~V.~Nanopoulos,
  ``No Scale Supersymmetric Guts,''
  Nucl.\ Phys.\  B {\bf 247} (1984) 373.
  
  J.~R.~Ellis, C.~Kounnas and D.~V.~Nanopoulos,
  ``Phenomenological SU(1,1) Supergravity,''
  Nucl.\ Phys.\  B {\bf 241} (1984) 406.

  J.~R.~Ellis, A.~B.~Lahanas, D.~V.~Nanopoulos and K.~Tamvakis,
 ``No-Scale Supersymmetric Standard Model,''
  Phys.\ Lett.\  B {\bf 134}, 429 (1984).

\bibitem{StringyNoscale} 
  E.~Witten,
  ``Dimensional reduction of superstring models,''
  Phys.\ Lett.\  B {\bf 155} (1985) 151.
 
  S.~Ferrara, C.~Kounnas and M.~Porrati,
  ``General dimensional reduction of ten-dimensional supergravity and
  superstring,''
  Phys.\ Lett.\  B {\bf 181} (1986) 263.
  
 M.~Cvetic, J.~Louis and B.~A.~Ovrut,
  ``A string calculation of the K\"ahler potentials for moduli of $\Z_N$
  orbifolds,''
  Phys.\ Lett.\  B {\bf 206} (1988) 227.
 
  L.~J.~Dixon, V.~Kaplunovsky and J.~Louis,
  ``On effective field theories describing $(2,2)$ vacua of the heterotic
  string,''
  Nucl.\ Phys.\  B {\bf 329} (1990) 27.
 
 M.~Cvetic, J.~Molera and B.~A.~Ovrut,
  ``K\"ahler potentials for matter scalars and moduli of $\Z_N$ orbifolds,''
  Phys.\ Rev.\  D {\bf 40} (1989) 1140.
  


\bibitem{WittenHoloFactor} 
  E.~Witten,
 ``Three-Dimensional Gravity Revisited,''
  arXiv:0706.3359 [hep-th].
 


\bibitem{HennStrominger}
  J.~D.~Brown and M.~Henneaux,
  ``Central Charges in the Canonical Realization of Asymptotic Symmetries: An
  Example from Three-Dimensional Gravity,''
  Commun.\ Math.\ Phys.\  {\bf 104}, 207 (1986).
 
  D.~Anninos, W.~Li, M.~Padi, W.~Song and A.~Strominger,
  ``Warped $AdS_3$ Black Holes,''
  arXiv:0807.3040 [hep-th].
  
  
   
\end{thebibliography}
\end{document}